\documentclass[twocolumn,letter]{jpsj3}

\usepackage{txfonts}
\usepackage{amsmath,amssymb,amsfonts,bm}
\usepackage{mathtools}
\usepackage{physics}
\usepackage{color}
\usepackage{ulem}

\title{Spatially Dispersive Second-Harmonic Generation in Ferroaxial Systems}

\author{Takumi Shiro and Satoru Hayami}
\inst{Graduate School of Science, Hokkaido University, Sapporo 060-0810, Japan}

\abst{
Spatially dispersive second-harmonic generation (SHG) provides a powerful probe of centrosymmetric multipolar states beyond the electric-dipole approximation. 
We develop a gauge-consistent microscopic theory of spatially dispersive SHG that treats electric-quadrupole (EQ) and magnetic-dipole (MD) processes on equal footing. 
Applying the formulation to a minimal triangular cluster model with ferroaxial order, which is closely related to an electric toroidal dipole, we show that the nonlinear optical response directly reflects the ferroaxial order parameter. 
The agreement between length- and velocity-gauge calculations confirms the gauge consistency of the formulation. 
We further demonstrate that the EQ and MD contributions exhibit different spectral weights despite sharing the same resonance energies. 
In particular, the MD channel can dominate over the EQ channel at selected resonances, indicating that it is not merely a perturbative correction. 
Our results establish that both EQ and MD processes are essential for a quantitative description of spatially dispersive SHG in ferroaxial materials and provide a microscopic basis for interpreting nonlinear optical signatures of centrosymmetric multipolar order.
}

\begin{document}
\maketitle

Nonlinear optical responses provide powerful probes of symmetry breaking and electronic structure in condensed matter systems~\cite{Kleinman_PhysRev.128.1761, shen1984principles, Sipe1987,Manfred2005,Hsieh_PhysRevLett.106.057401, Morimoto2016,Holder2020,Ni2022,Zhang2025_Universal}. 
Among them, second-harmonic generation (SHG) is one of the most sensitive probes of crystal symmetry~\cite{Manfred2005,butet2015optical,huang2024second}. 
Within the conventional electric-dipole (ED) approximation, it
allowed only in noncentrosymmetric systems and has therefore been widely employed to identify inversion-symmetry breaking. 
Beyond the ED approximation, finite wave-vector effects associated with spatial dispersion generate additional nonlinear optical responses through electric-quadrupole (EQ) and magnetic-dipole (MD) couplings \cite{Malashevich2010,Glazov2011,Cheng2017,Ahn2022,Gassner2023}. 
Such spatially dispersive responses are allowed even in centrosymmetric systems and provide direct access to unconventional multipolar degrees of freedom that remain inaccessible in conventional dipolar optical processes.

Among such multipolar states, ferroaxial order, which is closely related to an electric toroidal dipole~\cite{DUBOVIK1990145, kusunose2022generalization, hayami2024unified}, has recently attracted considerable attention because of its ability to generate unconventional transverse responses~\cite{cheong2021permutable,Hayami2022}.
Characterized by a spontaneous rotational distortion described by a macroscopic axial-vector order parameter~\cite{Hlinka_PhysRevLett.113.165502, Hlinka_PhysRevLett.116.177602}, ferroaxial order can induce a variety of off-diagonal transport and optical phenomena despite preserving both spatial inversion and time-reversal symmetries, such as spin-current generation~\cite{Roy_PhysRevMaterials.6.045004,Hayami2022}, antisymmetric thermopolarization~\cite{Nasu_PhysRevB.105.245125}, nonlinear transverse magnetization~\cite{inda2023nonlinear,du2026electric}, unconventional Hall effects~\cite{Hayami_PhysRevB.108.085124}, nonlinear magnetostriction~\cite{kirikoshi2023rotational}, nonlinear Edelstein effects~\cite{kirikoshi2026light}, and mutual conversion between polarity and chirality~\cite{cheong2019sos, cheong2021permutable, hayami2025chirality, Miki_PhysRevLett.134.226401, zhang2026interwined}. 
Understanding the optical signatures of ferroaxial order is therefore crucial for elucidating the microscopic origin of these anomalous responses. 
Since inversion symmetry remains preserved, conventional ED SHG is forbidden, making spatially dispersive SHG one of the most direct probes of ferroaxial order~\cite{Jin2020,Yokota2022,Sekine_PhysRevMaterials.8.064406}.

Despite this experimental progress, the microscopic understanding of spatially dispersive SHG remains incomplete. 
Theoretical studies of finite wave-vector nonlinear optical responses have long been hampered by difficulties associated with gauge invariance and spurious divergences. 
Recent gauge-covariant formulations have overcome these issues and established a consistent framework for nonlinear optical responses beyond the ED approximation~\cite{McKay2024,Zhang2025}.
In particular, Gassner and Mele developed a theory of first-order spatially dispersive SHG in crystalline systems that naturally incorporates both EQ and MD processes~\cite{Gassner2023}.
However, practical applications have focused almost exclusively on the EQ contribution, leaving the quantitative role of the MD channel largely unexplored.

In this Letter, we develop a gauge-consistent microscopic theory of spatially dispersive SHG that incorporates EQ and MD processes on equal footing.
Applying the formulation to a minimal ferroaxial cluster model~\cite{Hayami2023}, we uncover the microscopic origin of the ferroaxial nonlinear optical response.
We demonstrate that the relative importance of the EQ and MD channels depends sensitively on frequency and tensor component; whereas some resonances are predominantly EQ-driven, others are governed by the MD channel and cannot be captured within an EQ-only framework.
Our results reveal a nontrivial competition between EQ and MD processes and establish both channels as essential ingredients for a quantitative description of spatially dispersive SHG in ferroaxial materials.

To describe spatially dispersive SHG on a microscopic footing, we employ the density-matrix formalism for nonlinear optical responses~\cite{Aversa1995,Cabellos2009,Ventura2017,Passos2018,Parker2019,Watanabe2020,Watanabe2021}.
This approach provides a framework for treating EQ and MD processes on equal footing while maintaining gauge consistency.
The time evolution of an electronic system subjected to a time-dependent external field is described by the von Neumann equation.
\begin{align}
  i\hbar\partial_t\hat{\rho}(t)=[\hat{\mathcal{H}}(t),\hat{\rho}(t)],
\end{align}
where $\hat{\rho}(t)$ is the density matrix and $\hat{\mathcal{H}}(t)$ is the total Hamiltonian including the external field. 
To evaluate nonlinear optical responses, we decompose the total Hamiltonian into the unperturbed part $\hat{\mathcal{H}}_0$ and the perturbation part $\hat{V}(t)$, such that $\hat{\mathcal{H}}(t)=\hat{\mathcal{H}}_0+\hat{V}(t)$.

In the frequency domain, it is convenient to introduce the energy denominator
\begin{equation}
  d^\omega_{ab}=\frac{1}{\hbar\omega+i\eta-\epsilon_{ab}},
\end{equation}
where $\epsilon_{ab}=\epsilon_a-\epsilon_b$ and $\eta$ is a positive infinitesimal associated with the adiabatic switching-on of the external field.
Using the recursive formulation of the density matrix, the $(n+1)$th-order contribution can be written as
\begin{equation}
  \hat{\rho}^{n+1}_{\hat{V}_1 \cdots \hat{V}_{n+1}}=
  d^{\omega_\Sigma}\circ[\hat{V}_{n+1},\hat{\rho}^{n}_{\hat{V}_1 \cdots \hat{V}_n}],
\end{equation}
where the symbol $``\circ"$ denotes the Hadamard (element-wise) product of the matrices, and $\omega_\Sigma$ is the sum of the frequencies associated with the perturbations up to the $(n+1)$th order. 
Since SHG is a second-order optical response, we focus on the case $n=2$. The nonlinear current density at frequency 2$\omega$, denoted as $J_i(2\omega)$, is obtained from the expectation value of the current operator $\hat{j}_i$ for $i=x,y,z$:
\begin{equation}
  J_{i}(2\omega)=\text{Tr}[\hat{\rho}^{(2)}(2\omega)\hat{j}_i],
\end{equation}
which defines the second-order nonlinear optical conductivity tensor.
Within the ED approximation, optical responses are evaluated in the uniform limit $\mathbf q\rightarrow\mathbf0$.
Introducing the ED operator $\hat P_i=-er_i$, the current operator is given by $\hat{j}_i=\partial_t\hat{P}_i$.
To capture spatially dispersive SHG, however, it is necessary to retain the first-order contribution in the optical wave vector $\mathbf q$.
This finite wave-vector correction gives rise to EQ and MD couplings beyond the ED approximation and forms the basis of the spatially dispersive SHG studied in this work.

To describe higher-order multipolar responses beyond the ED approximation, it is necessary to retain the finite optical wave vector $\mathbf q$.
Expanding the light-matter coupling as $e^{i\mathbf q\cdot\mathbf r} \simeq 1+i\mathbf q\cdot\mathbf r$, the optical response can be organized in powers of $\mathbf q$. The zeroth-order term corresponds to the conventional ED response, whereas the linear-order correction gives rise to spatially dispersive SHG.
Accordingly, the macroscopic nonlinear current density at frequency $2\omega$ is expressed as
\begin{equation}
J_i(2\omega, \mathbf{q}) = \sigma_{ijk}^{(0)} E_j(\omega) E_k(\omega) + \sigma^{(1)}_{ijkl} q_j E_k(\omega) E_l(\omega),
\end{equation}
where $\sigma^{(0)}_{ijk}$ denotes the conventional SHG conductivity within the ED approximation and $\sigma^{(1)}_{ijkl}$ represents the first-order spatially dispersive contribution.
Since the ferroaxial phase preserves spatial inversion symmetry, the ED SHG tensor $\sigma^{(0)}_{ijk}$ vanishes identically, and the nonlinear response is governed entirely by the spatially dispersive term.
Following Refs.~\citen{Malashevich2010,Gassner2023}, the first-order spatially dispersive current operator can be decomposed into EQ and MD channels.
Introducing the EQ operator
\begin{equation*}
  \hat{Q}^{ji}=-er_jr_i,
\end{equation*}
and the MD operator
\begin{equation*}
  \hat{M}_k=-\frac{
  e}{2c}(\mathbf{r}\times\mathbf{\hat{j}})_k,
\end{equation*}
where $e$ is the elementary charge and $c$ is the speed of light.
Then, the current operator takes the form
\begin{equation}
  \hat{J}^i=\partial_t\hat{P}^{i}-iq_j
  \left( \frac{1}{2}\partial_t\hat{Q}^{ji}+c\epsilon^{jik}\hat{M}_k
  \right).
  \label{eq:current}
\end{equation}
where the first term describes the conventional ED current, while the finite wave-vector correction consists of EQ and MD contributions. 
The EQ term originates from charge redistribution associated with quadrupolar motion, whereas the MD term reflects circulating currents carrying orbital magnetization.

In the length gauge, the corresponding light-matter interaction is given by
\begin{equation}
  \hat{V}=
  \left[
  \hat{P}_i -q_j \left(\frac{i}{2} \hat{Q}^{ji}+\frac{c}{\omega}\epsilon^{jil}\hat{M}_l \right)
  \right] E_i.
  \label{eq:V}
\end{equation}
where $E_i$ is the electric field for $i=x,y,z$. This expression shows that both multipolar channels enter the optical response on an equal footing.
Although EQ and MD terms naturally appear in Eqs.~(\ref{eq:current}) and (\ref{eq:V}), previous microscopic studies of spatially dispersive SHG have focused primarily on the EQ contribution~\cite{Gassner2023}.
In contrast, the present work retains both channels throughout the calculation and evaluates their respective contributions to the ferroaxial SHG response.
  The resulting linear spatially dispersive SHG conductivity is expressed as
\begin{multline}
  \sigma^{(1)}_{ijkl}=
  \left(\frac{e^3}{2\hbar^2}
  \right)\int d\mathbf{k}\sum_{a,b,c}
  d^{2\omega}_{ba}\Big[
    \hat{v}^i_{ab}
    \big(
    \hat{r}^k_{bc}
    \hat{\mathcal{Q}}^{jl}_{ca}
    d^{\omega}_{ca}
    f_{ac}
    -
    \hat{r}^k_{ca}
    \hat{\mathcal{Q}}^{jl}_{bc}
    d^{\omega}_{bc}
    f_{cb}
    \big)\\
    +
    \hat{v}^i_{ab}
    \big(
    \hat{r}^l_{ca}
    \hat{\mathcal{Q}}^{jk}_{bc}
    d^{\omega}_{ca}
    f_{ac}
    -
    \hat{r}^l_{bc}
    \hat{\mathcal{Q}}^{jk}_{ca}
    d^{\omega}_{bc}
    f_{cb}
    \big)\\
    +i
    (\frac{1}{2}\partial_t\hat{Q}^{jl}_{ab}+c\epsilon_{jlm}\hat{M}^m_{ab})
    \big(
      \hat{r}^k_{bc}\hat{r}^l_{ca}d^{\omega}_{ca}f_{ac}
      -
      \hat{r}^l_{bc}\hat{r}^k_{ca}d^{\omega}_{bc}f_{cb}
    \big)\\
   +(k \leftrightarrow l)\Big].
\end{multline}

with
\begin{align}
\hat{\mathcal{Q}}^{jl}_{ab}\equiv \frac{i}{2}\hat{Q}^{jl}_{ab}+\frac{c}{\omega}\epsilon_{jlm}\hat{M}^m_{ab}, 
\end{align}

where $f_{ab} \equiv f(\epsilon_a)-f(\epsilon_b)$ is the difference in the Fermi--Dirac distribution functions between the Bloch states $a$ and $b$.
For later discussion, the conductivity is decomposed as
\[\sigma^{(1)}=\sigma_{\text{EQ}}+\sigma_{\text{MD}}.\]
This decomposition enables us to identify the microscopic origin of the ferroaxial SHG response in terms of distinct EQ and MD processes.
Hereafter, we omit the superscript (1) and denote the linear spatially dispersive SHG conductivity tensor simply by $\sigma_{ijkl}$.

For comparison with the length-gauge formulation, we also calculate the spatially dispersive SHG response in the velocity gauge~\cite{Gassner2023}.
The electromagnetic field is introduced via the minimal-coupling substitution $\mathbf{k} \rightarrow \mathbf{k} + e\mathbf{A}/\hbar$, where the vector potential is taken as  
$\mathbf{A}(\mathbf{r},\omega) = \mathbf{A}_{0}(\omega)e^{i\mathbf{q}\cdot\mathbf{r}}$.
Expanding the Hamiltonian with respect to $\mathbf{A}$ while retaining terms linear in the wave vector $\mathbf{q}$ gives the interaction Hamiltonian governing the spatially dispersive nonlinear optical response.

\begin{figure}[tb]
  \centering
  \includegraphics[width=0.5\linewidth]{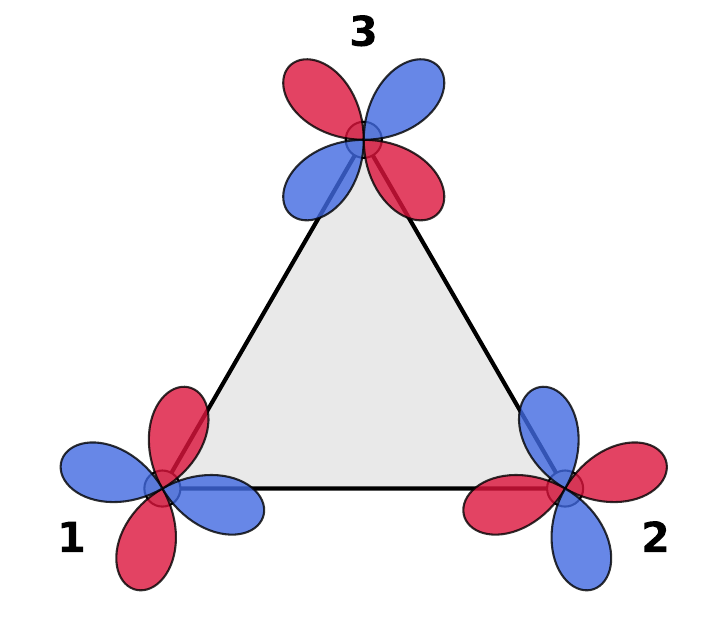}
  \caption{Schematic illustration of the triangular cluster model with ferroaxial order. The numbers 1, 2, and 3 indicate the cluster sites. The site-local quadrupoles
  form the $A_2^{\prime +}$ configuration, which induces the ferroaxial moment $G_z$ perpendicular to the cluster plane.}
  \label{fig:model}
\end{figure}

To examine spatially dispersive SHG responses in a ferroaxial system, we introduce a minimal triangular cluster model, as shown in Fig.~\ref{fig:model}.
The cluster lies in the $xy$ plane and consists of three sites connected by hopping processes.
In the absence of ferroaxial order, the system has $D_{\rm 3h}$ symmetry.
Ferroaxial order is introduced by a site-dependent arrangement of local electric quadrupoles, which breaks the in-plane mirror symmetries while preserving spatial inversion and time-reversal symmetries.
In the language of cluster multipoles~\cite{hayami2016emergent, Suzuki_PhysRevB.95.094406}, this quadrupole configuration corresponds to the axial-vector component $G_z$, i.e., the ferroaxial moment perpendicular to the cluster plane.
Consequently, the point-group symmetry is reduced from $D_{\rm 3h}$ to $C_{\rm 3h}$.

To describe this ferroaxial phase microscopically, we employ the tight-binding model~\cite{Hayami2023}.
The local Hilbert space is spanned by the two orbital basis states $\{\ket{d_{zx}},\ket{d_{yz}}\}$.
The Hamiltonian is written as
\begin{align}
  \hat{\mathcal{H}}_0&=\hat{\mathcal{H}}_t+\hat{\mathcal{H}}^Q_{\rm FA}\\
  \hat{\mathcal{H}}_t&=\sum_{ij\alpha\beta}(t^{\alpha\beta}_{ij}c^\dagger_{i\alpha} c_{j\beta}+\text{h.c.})\\
  \hat{\mathcal{H}}^Q_{\rm FA}&=-h_Q\sum_{i\alpha\beta}(\mathbf{e}_i\cdot\tilde{\mathbf{\tau}})^{\alpha\beta}c^\dagger_{i\alpha}c_{i\beta}.
\end{align}
Here, $c^\dagger_{i\alpha}(c_{i \alpha})$ creates (annihilates) an electron at site $i \in$\{1,2,3\} with orbital $\alpha \in$\{$zx, yz$\}.
The first term $\hat{\mathcal H}_t$ describes intersite hopping characterized by the Slater--Koster parameter $t_{dd\pi}=1.0$.
The second term $\hat{\mathcal H}^Q_{\rm FA}$ represents the ferroaxial mean field, where $h_Q$ controls the magnitude of the quadrupolar ordering.
Here, $\tilde{\mathbf{\tau}}$ denotes the Pauli matrices acting on the orbital subspace spanned by $\{d_{zx}, d_{yz}\}$, and $\mathbf{e}_i$ is the unit vector representing the local orientation of the quadrupole moment at site $i$, with $\mathbf{e}_1=(-\sqrt{3}/2,-1/2),\;  \mathbf{e}_2=(\sqrt{3}/2,-1/2)$, and
$\mathbf{e}_3=(0,1) $. 

The resulting $A_2^{\prime +}$ arrangement of local quadrupoles realizes the cluster multipole $G_z$ and provides the microscopic origin of the ferroaxial SHG response investigated below. 
It is noted that the ferroaxial degree of freedom does not exist on an individual atomic site in the cross-product form of the orbital and spin angular momenta~\cite{Hayami2022,Inda_PhysRevB.111.L041104} but emerges only through the collective arrangement of multipoles within the cluster.

Before presenting numerical results, we analyze the symmetry constraints on the spatially dispersive SHG tensor.
Using the irreducible representations of the parent point group $D_{\rm 3h}$, the optical electric field and the wave vector in the cluster plane both transform as the two-dimensional representation $E^\prime$.
Then, the in-plane tensor components obey
\begin{equation}
E^\prime\otimes E^\prime\otimes E^\prime\otimes E^\prime
=
2A_1^\prime\oplus 2A_2^\prime\oplus 4E^\prime.
\end{equation}

This decomposition shows that the in-plane spatially dispersive SHG tensor contains two independent $A_2^\prime$ components. 
These components are particularly important because they are activated by the ferroaxial symmetry lowering ($D_{\rm 3h}\to C_{\rm 3h}$) and therefore serve as direct optical probes of the ferroaxial order.

The nonzero components of the in-plane spatially dispersive SHG tensor $\sigma_{ijkl}$ are classified into the $A_1^{\prime}$ and $A_2^{\prime}$ sectors.
The $A_1^{\prime}$ sector is already allowed under the parent $D_{\rm 3h}$ symmetry and corresponds to fully symmetric multipoles such as ${Q_0,Q_{40}}$ in the multipole classification~\cite{Yatsushiro2021}. Its tensor components satisfy
\begin{equation}
\begin{aligned}
\sigma_{yyyy} &= \sigma_{xxxx}, \ \ \sigma_{yyxx} = \sigma_{xxyy}, \\
\sigma_{xyxy} &= \sigma_{xyyx} = \sigma_{yxxy} = \sigma_{yxyx} = \frac{\sigma_{xxxx} - \sigma_{xxyy}}{2}.
\end{aligned}
\end{equation}

In contrast,
the $A_2^{\prime}$ sector is induced by the ferroaxial order and is associated with the axial multipoles ${G_z,G_z^\alpha}$, where $G_z^\alpha$ is the electric toroidal octupole. 
These tensor components vanish in the parent $D_{\rm 3h}$ phase and become finite only after the symmetry is lowered to $C_{\rm 3h}$.
They therefore provide the symmetry-selective signature of ferroaxial order in spatially dispersive SHG. The corresponding relations are

\begin{equation}
\begin{aligned}
\sigma_{yxxx} &= -\sigma_{xyyy}, \ \  \sigma_{yxyy} = -\sigma_{xyxx}, \\
\sigma_{xxxy} &= \sigma_{xxyx} =-\sigma_{yyxy} = -\sigma_{yyyx} = \frac{\sigma_{xyyy} - \sigma_{xyxx}}{2}.
\end{aligned}
\end{equation}

Thus, the ferroaxial SHG response can be identified by focusing on the $A_2^{\prime}$ tensor components, such as $\sigma_{xyyy}$ and $\sigma_{xyxx}$.

\begin{figure}[tb]
  \centering
  \includegraphics[width=0.8\linewidth]{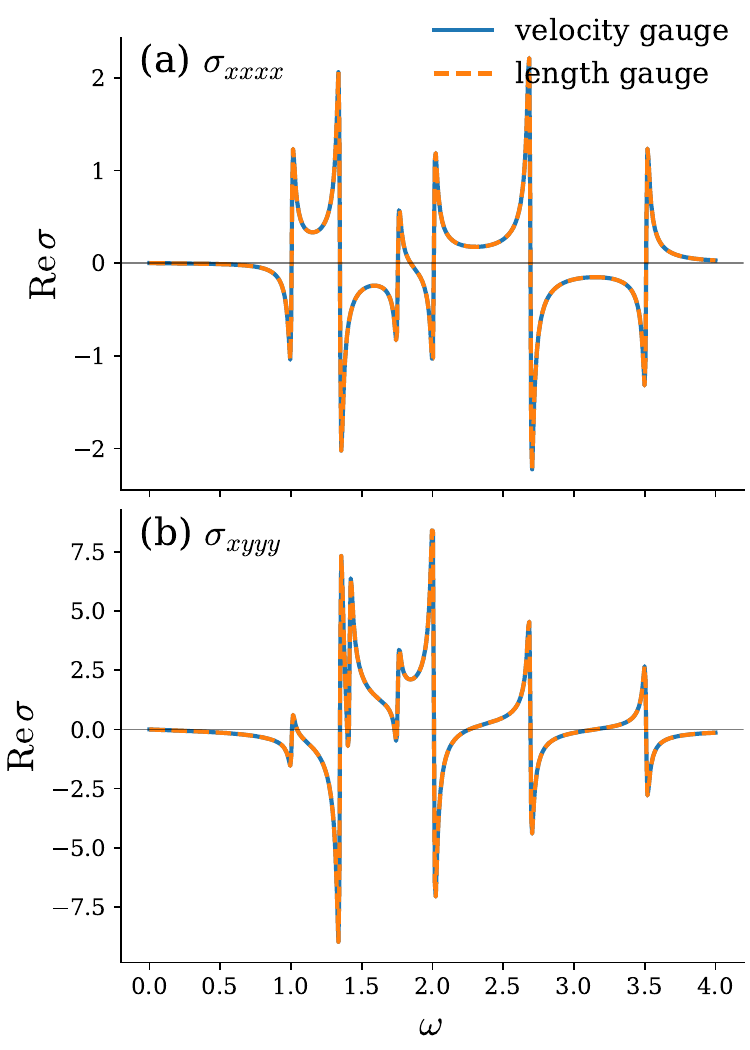}
  \caption{Comparison between the length- and velocity-gauge calculations for the real parts of the spatially dispersive SHG conductivity: (a) $\sigma_{xxxx}$ and (b) $\sigma_{xyyy}$. 
  Solid and dashed lines represent the velocity- and length-gauge results, respectively.
  }
  \label{fig:gauge}
\end{figure}

Before analyzing the EQ and MD contributions, we first verify the gauge consistency of the formulation.
We adopt the lattice basis $\mathbf{e}_i$, defining the position operator $\hat{\mathbf{r}}$ and the velocity operator $\hat{\mathbf{v}}=(i/\hbar)[\hat{\mathcal{H}}_0,\hat{\mathbf{r}}]$.
Figure~\ref{fig:gauge} compares the spatially dispersive SHG conductivities obtained in the length and velocity gauges for two representative tensor components.
Unless otherwise stated, we set $h_Q=1$, $\hbar=1$, and $\eta=0.01$, and consider a half-filled system at zero temperature. 
Figure~\ref{fig:gauge}(a) shows an $A_1^\prime$ component that is already allowed by the parent $D_{\rm 3h}$ symmetry, whereas Fig.~\ref{fig:gauge}(b) shows an $A_2^\prime$ component associated with the ferroaxial order and activated by the symmetry lowering $D_{\rm 3h}\rightarrow C_{\rm 3h}$.
In both cases, the length- and velocity-gauge calculations agree well over the entire frequency range, confirming that the present implementation consistently describes the finite-wave-vector nonlinear response.

\begin{figure}[tb]
  \centering
  \includegraphics[width=0.9\linewidth]{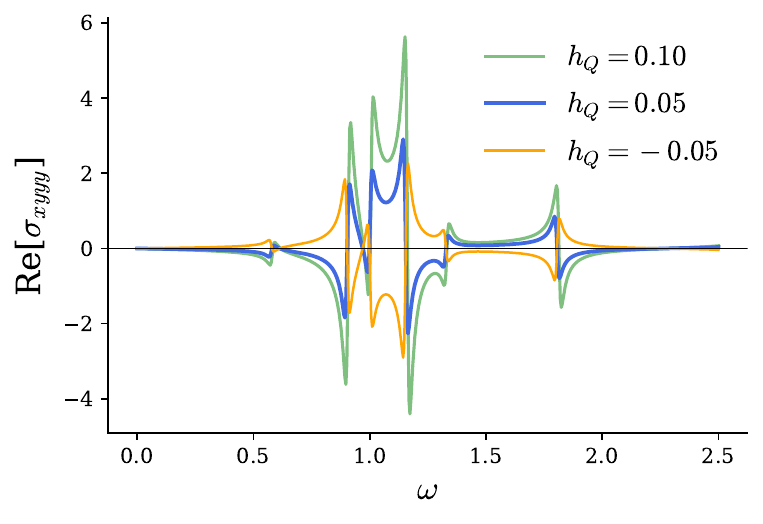}
  \caption{
  Frequency dependence of $\sigma_{xyyy}$ for several values of the ferroaxial mean field $h_Q$.
  }
  \label{fig:hq}
\end{figure}

We next focus on the ferroaxial tensor component $\sigma_{xyyy}$ belonging to the $A_2^\prime$ sector.
Figure~\ref{fig:hq} shows its frequency dependence for several values of the ferroaxial mean field $h_Q$.
The response reverses sign under $h_Q \rightarrow -h_Q$ and increases in magnitude with increasing $|h_Q|$.
Moreover, in the weak mean-field regime, the conductivity is approximately proportional to $h_Q$.
These features demonstrate that $\sigma_{xyyy}$ couples directly to the ferroaxial order parameter.
This behavior is consistent with the symmetry analysis, since $\sigma_{xyyy}$ is activated only by the ferroaxial symmetry lowering $D_{\rm 3h}\rightarrow C_{\rm 3h}$ and therefore provides a direct optical signature of the ferroaxial moment $G_z$.

\begin{figure}[tb]
  \centering
  \includegraphics[width=0.9\linewidth]{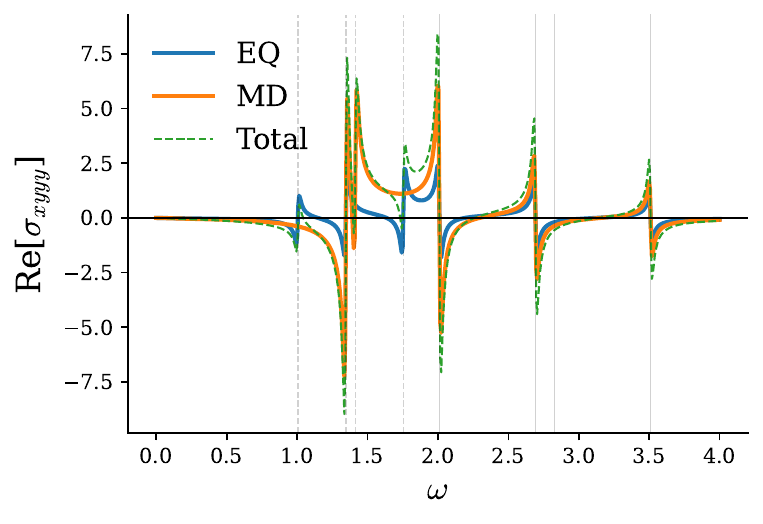}
  \caption{Frequency dependence of the EQ, MD, and total contributions to the ferroaxial SHG conductivity ${\rm Re} [\sigma_{xyyy}]$.
  The vertical and solid gray lines denote the resonance energies associated with the $\omega$ and 2$\omega$ denominators, respectively.
  }
  \label{fig:EQ_MD}
\end{figure}

We then investigate its microscopic origin by decomposing the SHG conductivity into the EQ and MD contributions.
Figure~\ref{fig:EQ_MD} shows the frequency dependence of ${\rm Re}[\sigma_{xyyy}]$ together with its EQ, MD, and total components.

Although the two channels exhibit resonances at similar frequencies, their spectral weights differ substantially.
Several resonances are selectively enhanced in only one channel, indicating that the EQ and MD operators probe distinct aspects of the optical transitions through different matrix elements and selection rules.

A particularly clear example appears around $\omega\simeq1.4$, where the MD contribution dominates while the EQ response remains weak.
This result demonstrates that the MD channel cannot generally be regarded as a perturbative correction to the EQ response.
Instead, it can provide a qualitatively distinct contribution and even become the dominant source of the ferroaxial SHG signal at specific resonances.

These results reveal that the ferroaxial SHG response originates from a nontrivial interplay between EQ and MD processes.
Consequently, a description based solely on the EQ channel may overlook characteristic spectral features associated with ferroaxial order, highlighting the necessity of treating both channels on equal footing for a quantitative understanding of spatially dispersive SHG.

To summarize, we have investigated spatially dispersive SHG in a ferroaxial triangular cluster model by treating EQ and MD processes on an equal footing.
The agreement between the length- and velocity-gauge calculations confirms the gauge consistency of the present formulation.
The sign reversal and nearly linear dependence on the ferroaxial mean field $h_Q$ demonstrate that the calculated $A_2^\prime$ tensor component directly reflects the ferroaxial order parameter.

By decomposing the ferroaxial SHG conductivity into EQ and MD contributions, we found that the two channels exhibit qualitatively different spectral characteristics.
Although their resonance energies are governed by the same electronic excitation spectrum, their spectral weights strongly depend on frequency and tensor component.
In particular, several resonances are dominated by the MD channel, indicating that the MD response is not merely a small correction to the EQ response but can constitute an essential part of the ferroaxial nonlinear optical signal.
These results establish spatially dispersive SHG as a symmetry-selective probe of ferroaxial order and demonstrate the importance of treating EQ and MD processes on an equal footing.
The present results provide a microscopic basis for interpreting SHG experiments in centrosymmetric ferroaxial materials. 
Since conventional ED SHG is forbidden in such systems, observed SHG signals necessarily originate from higher-order multipolar processes, including EQ and MD channels.

Our findings indicate that the relative importance of these two channels should be carefully considered when analyzing SHG experiments on ferroaxial materials, such as RbFe(MoO$_4$)$_2$~\cite{Jin2020, Hayashida_PhysRevMaterials.5.124409}, NiTiO$_3$~\cite{hayashida2020visualization, Hayashida_PhysRevMaterials.5.124409, Yokota2022, Bhowal_PhysRevResearch.6.043141}, and MnTiO$_3$~\cite{Sekine_PhysRevMaterials.8.064406}, where ferroaxial domains or symmetry breaking have been identified by SHG measurements.
The present framework is also applicable to a wide range of ferroaxial materials, such as Cu$_3$Nb$_2$O$_8$~\cite{Johnson_PhysRevLett.107.137205}, CaMn$_7$O$_{12}$~\cite{Johnson_PhysRevLett.108.067201}, BaCoSiO$_4$~\cite{Xu_PhysRevB.105.184407}, Ca$_5$Ir$_3$O$_{12}$~\cite{Hasegawa_doi:10.7566/JPSJ.89.054602, hanate2021first, hanate2023space}, K$_2$Zr(PO$_4$)$_2$~\cite{yamagishi2023ferroaxial, Bhowal_PhysRevResearch.6.043141, xie2026spinless}, Na$_2$Hf(BO$_3$)$_2$~\cite{nagai2023chemicalSwitching}, Na-superionic conductors~\cite{nagai2023chemical}, and 1$T$-TaS$_2$~\cite{Luo_PhysRevLett.127.126401, liu2023electrical}.
Extending the present framework to realistic models will be an important step toward identifying material-specific fingerprints of ferroaxial order in nonlinear optical spectra.

\begin{acknowledgment}
This research was supported by JSPS KAKENHI Grants Numbers JP22H00101, JP23H04869, and by JST CREST (JPMJCR23O4) and JST FOREST (JPMJFR2366).
\end{acknowledgment}

\bibliographystyle{jpsj}
\bibliography{reference}

\end{document}